\documentclass[10pt,conference]{IEEEtran}
\IEEEoverridecommandlockouts
\usepackage{cite}
\usepackage{amsmath,amssymb,amsfonts,bm}
\usepackage{algorithmic}
\usepackage{graphicx}
\usepackage{subcaption}
\usepackage{textcomp}
\usepackage{xcolor}
\usepackage{bbold}
\usepackage{mwe}

\usepackage[ruled,vlined]{algorithm2e}
\usepackage{array}
\usepackage{booktabs}

\def\BibTeX{{\rm B\kern-.05em{\sc i\kern-.025em b}\kern-.08em
    T\kern-.1667em\lower.7ex\hbox{E}\kern-.125emX}}
    
    \usepackage{amsfonts}

\newcommand{\overbar}[1]{\mkern 1.5mu\overline{\mkern-1.5mu#1\mkern-1.5mu}\mkern 1.5mu}

\begin{document}

\title{Consensus-based Distributed Variational Multi-object Tracker in Multi-Sensor Network}

\author{\IEEEauthorblockN{Qing Li, Runze Gan, Simon Godsill}
\IEEEauthorblockA{\textit{Engineering Department} \\
\textit{University of Cambridge}\\
Cambridge, UK \\
\{ql289, rg605, sjg30\}@cam.ac.uk}
}
\maketitle
\begingroup\renewcommand\thefootnote{\textsection}

\begin{abstract}
The growing need for accurate and reliable tracking systems has driven significant progress in sensor fusion and object tracking techniques. In this paper, we design two variational Bayesian trackers that effectively track multiple targets in cluttered environments within a sensor network. We first present a centralised sensor fusion scheme, which involves transmitting sensor data to a fusion center. Then, we develop a distributed version leveraging the average consensus algorithm, which is theoretically equivalent to the centralised sensor fusion tracker and requires only local message passing with neighbouring sensors. In addition, we empirically verify that our proposed distributed variational tracker performs on par with the centralised version with equal tracking accuracy. Simulation results show that our distributed multi-target tracker outperforms the suboptimal distributed sensor fusion strategy that fuses each sensor's posterior based on arithmetic sensor fusion and an average consensus strategy.
\end{abstract}


\section{Introduction} \label{sec:intro}
The distributed multi-sensor multi-object tracker has emerged as a promising approach due to its potential for reduced communication costs and increased robustness against single-node faults when compared to centralised fusion solutions. Several optimal algorithms for distributed data fusion have been developed, relying solely on local message passing \cite{chong1990distributed,chong2017forty}. However, these techniques necessitate specific network topologies, such as fully connected and tree-connected networks, and often come with a high computational burden that limits their applicability in certain situations. 

To overcome the limitations, several approximate methods have been studied. One popular fusion strategy is geometric average fusion, such as the widely used generalised covariance intersection method proposed in \cite{mahler2000optimal}, with the aim of avoiding double counting of common information while fusing multiple multi-object densities with unknown correlations among sensors. An alternative approach is the arithmetic average fusion, which has been shown to perform better when fusing random variables or point estimates \cite{kayaalp2022arithmetic}. Consensus-based algorithms \cite{olfati2004consensus,xiao2004fast,xiao2005scheme} have been introduced to enable geometric or arithmetic average fusion in a fully distributed manner. The analysis and comparison of these two fusion strategies can be found in \cite{li2021distributed,kayaalp2022arithmetic}.
However, these methods that fuse the local posteriors of each sensor are suboptimal and can result in degraded tracking performance. In \cite{hlinka2012likelihood}, a consensus-based method was designed to obtain an approximation of the joint likelihood function by distributing the likelihood functions of each sensor. This likelihood consensus method was then developed to implement distributed particle filters and distributed Gaussian particle filters for multiple target tracking applications. Nevertheless, the joint likelihood function is approximate, and the estimation accuracy and fusion efficiency can be affected by the choice of the basis functions.

Here we propose a solution for distributed sensor fusion and object tracking by leveraging the non-homogeneous Poisson process (NHPP) measurement model and the recently-developed NHPP trackers \cite{gan2022variational}. The original NHPP tracker in \cite{gilholm2005poisson} successfully avoids the data association problem, but its particle filter implementation is limited by the curse of dimensionality. To address this issue, an association-based NHPP measurement model was introduced in \cite{li2022scalable} to enable efficient parallel computing and a tractable structure. Additionally, a fast Rao-Blackwellised sequential Markov chain Monte Carlo sampling scheme was developed in \cite{li2023adaptive} with improved efficiency compared to \cite{li2022scalable} for linear Gaussian models. While sampling-based methods like those presented in \cite{li2022scalable,li2023adaptive,li2021sequential} can theoretically converge to optimal Bayesian filters, their computational requirements can be intensive when the number of targets and measurements increases. Therefore, a high-performance variational inference implementation was designed in \cite{gan2022variational}, which achieves comparable tracking accuracy with sampling-based implementations \cite{li2022scalable} while offering faster processing speeds. 

This paper develops an extension of the variational Bayes multi-object tracker presented in \cite{gan2022variational} to multi-sensor cases, as it has demonstrated superior tracking performance in terms of both accuracy and implementation efficiency. Our key contribution is the development of a variational filtering framework for tracking multiple objects in a distributed sensor network. This is accomplished by leveraging the average consensus algorithm, which, when successfully converged, allows the distributed version to be theoretically equivalent to the centralised sensor fusion tracker. In particular, each sensor in the network runs locally using its own measurements while communicating with its neighbouring sensors to obtain global statistics for the local coordinate ascent update. Once the average consensus algorithm has converged, the local estimates for each sensor are updated using the global statistics obtained from the consensus. Therefore, this approach only requires communication with neighbouring sensors and does not require complete knowledge of the network topology. Overall, the proposed approach allows for distributed sensor fusion and tracking that can achieve tracking accuracy equivalent to centralised fusion while being more efficient in communication costs. The simulation results show that compared to the arithmetic fusion method that fuses the local posteriors of each sensor, the proposed distributed variational tracker exhibits superior tracking accuracy and efficiency.

\section{Problem Formulation and Modelling} \label{sec:PoissonModel}
This paper considers tracking multiple targets in clutter under a distributed sensor network where the communication links between sensors can be time-varying. Assume that there are $K$ targets in the surveillance area. At each discrete time step $n$, their joint state is $X_n=[X_{n,1}^\top,X_{n,2}^\top,...,X_{n,K}^\top]^\top$, where each vector $X_{n,k}, k\in \{1,...,K\}$ denotes the kinematic state for the $k$-th target.
Suppose that the targets are observed by a sensor network consisting of $N_s$ sensors, each capable of observing the entire tracking area. The time-varying sensor network at time $t$ can be modelled as a graph $G(t) = \{\mathcal{S}, \mathcal{E}(t)\}$ at any given continuous time $t$, where the sensor set is denoted by $ \mathcal{S} = \{1, 2, \ldots, N_s\}$, and $\mathcal{E}(t)$ is the set of edges with the existence of edge $(i, j)$ meaning that the $i$-th sensor can communicate with the $j$-th sensor at time $t$. The set of neighbours of sensor $i$ is denoted by $\mathcal{N}_i(t) = \{j \mid (i, j) \in \mathcal{E}(t)\}$. The degree $d_i(t)$ of the $i$-th sensor represents the number of its neighbouring sensors with which it can communicate, i.e., $d_i(t) = |\mathcal{N}_i(t)|$. In a sensor network, the measurements received from all sensors at time step $n$ can be denoted by $Y_n=[Y_{n}^{1},Y_{n}^{2},...,Y_{n}^{N_s}]$. Each $Y_{n}^{s}$ includes measurements acquired by the $s$-th sensor, and $Y_{n}^{s}=[Y_{n,1}^{s},...,Y_{n,M_n^{s}}^{s}]$, where $M_n^{s}$ is the total number of measurements received at the $s$-th sensor ($s =1,..., N_s$). Subsequently, $M_n=[M_n^1,...,M_n^{N_s}]$ records the total number of measurements received from all sensors at time step $n$.
\vspace{-0.5em}
\subsection{Dynamical  model}\label{dynamic model}
We assume that targets move in a 2D surveillance area with each $X_{n,k}=[x^1_{n,k}, \Dot{x}^1_{n,k},x^2_{n,k}, \Dot{x}^2_{n,k}]^T$, where $x^d_{n,k}$ and $\Dot{x}^d_{n,k}$ ($d=1,2$) indicate the $k$-th target's position and velocity in the $d$-th dimension, respectively. We assume an independent linear Gaussian transition density for each target's states:
\vspace{-0.7em}
\begin{equation}\label{eq: dynamic transition}
    p(X_n|X_{n-1})
    =\prod_{k=1}^K\mathcal{N}(X_{n,k};F_{n,k}X_{n-1,k},Q_{n,k}). \vspace{-0.7em}
\end{equation}
where $F_{n,k}=diag(F^1_{n,k},F^2_{n,k})$, $Q_{n,k}=diag(Q^1_{n,k},Q^2_{n,k})$. For a constant velocity (CV) model, $F_{n,k}^d,Q_{n,k}^d$ ($d=1,2$) are
\vspace{-0.5em}
\begin{equation} \label{eq:model para}
    F_{n,k}^d=\begin{bmatrix} 1 & \tau \\ 0& 1
    \end{bmatrix}, 
    Q_{n,k}^d=\sigma_k^2\begin{bmatrix} \tau^3/3 & \tau^2/2 \\ \tau^2/2& \tau
    \end{bmatrix}, 
\end{equation}
where $\tau$ is the time interval between time steps.

\subsection{NHPP measurement model and association prior}
Here, we assume each sensor independently detects targets in accordance with the NHPP measurement model described in \cite{gilholm2005poisson}. Notably, the NHPP model may vary for each sensor.
Denote the set of Poisson rates for all sensors as $\Lambda=[\Lambda^{1},\Lambda^{2},...,\Lambda^{N_s}]$. For each sensor $s$, the Poisson rate vector is defined by $\Lambda^{s}=[\Lambda_0^{s},\Lambda_1^{s},...,\Lambda_K^{s}]$, where $\Lambda_0^{s}$ is the clutter rate and $\Lambda_k^{s}$ is the $k$-th target rate, $k=1,...,K$. For each sensor $s$, each target $k$ generates measurements by a NHPP with a Poisson rate $\Lambda_k^{s}$, and the total measurement process is also a NHPP from the superposition of the conditional independent NHPP measurement process from $K$ targets and clutter. The total number of measurements from the $s$-th sensor follows a Poisson distribution with a rate of $\Lambda_{sum}^{s}=\sum_{k=0}^K\Lambda_k^{s}$. 

Our independent measurement model assumption signifies that given $X_{n}$, the measurements of each sensor are conditionally independent, i.e., $p(Y_n|X_n)=\prod_{s=1}^{N_s} p(Y_{n}^{s}|X_{n})$.
We denote the associations of all measurements $Y_n$ by $\theta_{n}=[\theta_{n}^{1},\theta_{n}^{2},...,\theta_{n}^{N_s}]$
, with each $\theta_{n}^{s}=[\theta_{n,1}^{s},\theta_{n,2}^{s},...,\theta_{n,M_n^{s}}^{s}]$ ($s=1,...,N_s$) representing the association vector for the $s$-th sensor's measurements. Each component $\theta_{n,j}^{s}$ ($j=1,...,M_n^{s}$) gives the origin of the measurement $Y_{n,j}^{s}$; $\theta_{n,j}^{s}=0$ indicates that $Y_{n,j}^{s}$ is generated by clutter, and $\theta_{n,j}^{s}=k$ ($k=1,...,K$) means that $Y_{n,j}^{s}$ is generated from the target $k$. The adopted conditionally independent NHPP measurement model leads to the following properties: the joint association prior are conditionally independent give all the measurement numbers
\begin{equation}\label{eq:joint association prior}
    p(\theta_n|M_n) =\prod_{s=1}^{N_s}  p(\theta_{n}^{s}|M_n^{s}),
\end{equation}
and given all associations, the joint likelihood $p(Y_n|\theta_n,X_n)$ remains conditionally independent for each sensor, i.e.,
\begin{equation}\label{eq:conditional distribution of Y_n}
    p(Y_n|\theta_n,X_n)=\prod_{s=1}^{N_s} p(Y_{n}^{s}|\theta_{n}^{s},X_{n}).
\end{equation}

Finally, for each sensor $s$, the NHPP measurement model implies the following according to \cite{gan2022variational}:
measurements are conditionally independent given associations and target states
\begin{align}
    \label{eq:obs prior}
p(Y_{n}^{s}|\theta_{n}^{s},X_{n})&=\prod_{j=1}^{M_n^{s}}\ell^s(Y_{n,j}^{s}|X_{n,\theta_{n,j}^{s}}),
\end{align}
where $M_n^{s}$ is implicitly known from $\theta_n^{s}$ since $M_n^{s}=|\theta_n^{s}|$, and $\ell^s$ is the probability density function of a single measurements received in sensor $s$ given its originator's state. Here we assume the target originated measurement follows a linear and Gaussian model while the clutter measurement is uniformly distributed in the observation area of volume $V^{s}$:
\begin{equation} \ell^s(Y_{n,j}^{s}|X_{n,k})=\begin{cases} 
    \mathcal{N}(H X_{n,k},R_{k}^{s}),& \text{$k\neq 0 $; \ \ (object)}\\
     \frac{1}{V^{s}}, & \text{$k= 0 $; \ \ (clutter)}
\end{cases}
\label{measurement model}
\end{equation}
where $H$ is the observation matrix, and $R_{k}^{s}$ indicates the $s$-th sensor noise covariance. Moreover, the joint prior $p(\theta_{n}^{s}|M_n^{s})$ can be factorised as the product of $M_n^{s}$ independent association priors, i.e., $p(\theta_{n}^{s}|M_n^{s})=\prod_{j=1}^{M_n^{s}} p(\theta_{n,j}^{s})$,
where the prior for each association $p(\theta_{n,j}^{s})$ is a categorical distribution with support $\theta_{n,j}^{s} \in \{0,...,K\}$
\vspace{-0.7em}
\begin{align}
    \label{eq:single assoc prior}    &p(\theta_{n,j}^{s})=\frac{\sum_{k=0}^K\Lambda_k^{s}\delta[\theta_{n,j}^{s}=k]}{\Lambda_{sum}^{s}}.
\end{align}

\section{Coordinate Ascent Variational Filtering for Centralised Sensor Fusion} \label{sec: cavf}
This section develops a coordinate ascent variational filtering framework for tracking multiple objects in clutter in a centralised sensor network where there exists a central hub for collecting the measurements from  multiple sensors and using them to track the targets. The parameters $K,\Lambda,$ and $R_{1:K}^s$ in Section \ref{sec:PoissonModel} are assumed to be known and are therefore always implicitly conditioned in our derivations.
The objective is to sequentially estimate the posterior $p(X_n,\theta_n|Y_{1:n})$ given observations ${Y}_{1:n}$ from all sensors in the network. Accordingly, the exact optimal filtering can be recursively expressed as follows,
\vspace{-1em}
\begin{align} \label{eq:general exact filter with para learn}
    p(X_n,\theta_n|{Y}_{1:n})&\propto
    p(Y_n|\theta_n,X_n) p(\theta_n|M_n)  \\
    &~~\times \int p(X_n|X_{n-1}) p({X}_{n-1}|{Y}_{1:n-1})d{X}_{n-1},\notag
\end{align}
However, this exact filtering recursion is intractable, prompting us to replace $p({X}_{n-1}|{Y}_{1:n-1})$ with a tractable approximate filtering prior. According to \cite{gan2022variational}, a natural choice of this tractable prior is the approximate filtering result/posterior from the previous time step. In our context, where variational Bayes is employed to approximate the target distribution, this corresponds to using the converged variational distribution $q^*_{n-1}({X}_{n-1})$ obtained from the approximate filtering at time step $n-1$.
Therefore, the target posterior distribution of our current approximate filtering step is  
\vspace{-0.2em}
\begin{equation}  \label{eq:phatjoint}
    \hat{p}_n(X_{n},\theta_{n}|Y_n)  
    \propto p(Y_n|\theta_n,X_n)p(\theta_n|M_n)\hat{p}_n(X_n),
    \vspace{-0.2em}
\end{equation}
where the predictive prior $\hat{p}_n(X_n)$ is written as
\vspace{-0.2em}
\begin{equation}  \label{eq: predictive prior}
     \hat{p}_n(X_n) = \int p(X_n|X_{n-1})q^*_{n-1}(X_{n-1})dX_{n-1}. \vspace{-0.2em}
\end{equation}



\subsection{Coordinate ascent update} \label{sec:ca update}
We assume a mean-field family of variational distributions that satisfy the factorisation $q_n(X_n,\theta_n)=q_n(X_n)q_n(\theta_n)$. Then, the variational distribution $q^*_n(X_n,\theta_n)$ is chosen from the posited family that minimises the KL divergence $\text{KL}(q_n(X_n)q_n(\theta_n)||\hat{p}_n(X_n,\theta_n|Y_n))$.
This optimisation
with respect to $q_n$ can be done by the following coordinate ascent algorithm that ensures convergence.
We start by setting the initial association distribution $q_{n}(\theta_n)$ as $q_{n}^{(0)}(\theta_n)$;
afterwards, we iteratively update $q_n(X_n)$ while keeping $q_n(\theta_n)$ fixed, and update  $q_n(\theta_n)$ while keeping $q_n(X_n)$ fixed, repeating these steps until convergence is achieved. The converged variational distribution $q^*_n(X_{n},\theta_{n})$ is then used to approximate the target distribution $\hat{p}_n(X_{n},\theta_{n}|{Y}_n)$. We now present these updates.

\subsubsection{update for $q_n(X_n)$} \label{sec:ca update for Xn} 
First we present the update for $X_n$
\vspace{-0.8em}
\begin{equation} \label{eq: X update}
    q_n(X_n)
    \propto \hat{p}_n(X_n)\prod_{k=1}^{K}\mathcal{N}\left(\overbar{Y}_n^k;HX_{n,k},\overbar{R}^k_n\right), \vspace{-0.7em}
\end{equation}
where
\vspace{-0.7em}
\begin{equation} \label{eq:pseudomeas Y}
\overbar{R}_n^k={\left(\sum_{s=1}^{N_s} \Omega_{k,1}^{s}  \right)}^{-1}, ~~~~ \Omega_{k,1}^{s}=(R_k^{s})^{-1}\sum_{j=1}^{M_n^{s}}q_n(\theta_{n,j}^{s}=k), \vspace{-0.3em}
\end{equation}
\begin{equation} \notag
\overbar{Y}_n^k=\overbar{R}_n^k \sum_{s=1}^{N_s} \Omega_{k,2}^{s},~~~~\Omega_{k,2}^{s}=(R_k^{s})^{-1}\sum_{j=1}^{M_n^{s}}q_n(\theta_{n,j}^{s}=k)Y_{n,j}^{s}.
\end{equation}
Such an update can be considered as updating the predictive prior $\hat{p}_n(X_n)$ in \eqref{eq: predictive prior} with $K$ pseudo-measurements $\overbar{Y}_{n}^k,k=1,2,...,K$. Given an independent initial Gaussian prior $p(X_0)=\prod_{k=1}^K p(X_{0,k})$ and the transition in \eqref{eq: dynamic transition}, the updated variational distribution can always be in an independent Gaussian form, i.e., $q_n(X_n)=\prod_{k=1}^K q_n(X_{n,k})$.
Denote the converged variational distribution for the $k$-th target at time step $n-1$ as $q^*_{n-1}(X_{n-1,k})= \mathcal{N}(X_{n-1,k};\mu^{k*}_{n-1|n-1},\Sigma^{k*}_{n-1|n-1})$, then we denote its predictive prior according to \eqref{eq: predictive prior} by
\begin{align}
    \begin{aligned} \label{eq:predictive prior computation}
    \hat{p}_n(X_{n,k})=&\mathcal{N}(X_{n,k};\mu^{k*}_{n|n-1},\Sigma^{k*}_{n|n-1}).
    \end{aligned}
\end{align}
The variational distribution $q_n(X_{n,k})=\mathcal{N}(X_{n,k};\mu^{k}_{n|n},\Sigma^{k}_{n|n})$ can then be updated by Kalman filtering.
Such an update can be independently carried out for all targets.
\subsubsection{update for $q_n(\theta_n)$}
Since $q_n(\theta_n)=\prod_{s=1}^{N_s}  q_n(\theta_n^{s})$, $q_n(\theta_n)$ can be updated by individually evaluating $q_n(\theta_n^{s})$ for each sensor, where the update can be performed in parallel:
\begin{align} 
   & q_n(\theta_n^{s})   
    \propto\prod_{j=1}^{M_n^{s}}q_n(\theta_{n,j}^{s}),\\  \label{eq:update theta}
   & q_n(\theta_{n,j}^{s})
    \propto\frac{\Lambda_0^{s}}{V^{s}}\delta[\theta_{n,j}^{s}=0]+\sum_{k=1}^K\Lambda_k^{s} l_k^{s}\delta[\theta_{n,j}^{s}=k],\\  \notag
   & l_k^{s}=\mathcal{N}(Y_{n,j}^{s};H\mu_{n|n}^k,R_k^{s})\text{exp}(-0.5\text{Tr}({(R_k^{s})}^{-1}H\Sigma_{n|n}^k H^\top)),
\end{align}
where each $q_n(\theta_{n,j}^{s})$ is a categorical distribution and the updates for $\theta_n$ can be independently carried out for each $\theta_{n,j}$.

\subsection{Initialisation}
We adopt the initialisation strategy in \cite{gan2022variational}: at time step $n$, the algorithm starts the recursive updates from $q_n(X_n)$ and the initial variational distribution $q_n^{(0)}(\theta_n^{s})$ for each sensor $s$ is 
\begin{align} 
\label{eq:init independent theta}
&q_n^{(0)}(\theta_n^{s})
    \propto\frac{\Lambda_{0}^{s}}{V^{s}}\delta[\theta_{n,j}^{s}=0]+\sum_{k=1}^K\Lambda_k^{s} l_k^{s,0}\delta[\theta_{n,j}^{s}=k],\\ \notag 
   & l_k^{s,0}=\mathcal{N}(Y_{n,j}^{s};H\mu^{k*}_{n|n-1},H\Sigma^{k*}_{n|n-1}H^\top+R_k^{s}),
\end{align}

\begin{algorithm} 
\SetAlgoLined
\textbf{Input}: $q^*_{n-1}(X_{n-1}),Y_n,M_n$, maximum iteration $I_{max}$.\\
\textbf{Initialisation}: 
Set $\hat{p}_n(X_n)$ according to \eqref{eq:predictive prior computation}.\\
\textbf{At each sensor $s$:}\\
Initialise $q_n(\theta_{n}^{s})$ according to \eqref{eq:init independent theta}.\\
   \For {$i=1,2,...,I_{max}$}  
   {
   Compute $\Omega_{k,1}^{s}$, $\Omega_{k,2}^{s}$, $k=1,2,...,K$ using \eqref{eq:pseudomeas Y}.\\
   \textbf{Perform average consensus} with $\hat{\Omega}_{k,1}$ and $\hat{\Omega}_{k,2}$.\\
   \For {$k=1,2,...,K$}
   { 
   Evaluate $\overbar{R}_n^k,\overbar{Y}_n^k$ according to \eqref{eq:average consesus result}.\\
   Update $q_n(X_{n,k})$ by Kalman filtering.
   }
    
   Update $q_n(\theta_{n,j}^{s})$, $j=1,2,...,M_n^{s}$ using \eqref{eq:update theta}. \\

    }
   
   Set $q^*_{n}(X_{n})=\prod_{k=1}^Kq_n(X_{n,k})$, and $q^*_{n}(\theta_{n}^{s})=\prod_{j=1}^{M_n^{s}}q_n(\theta_{n,j}^{s}).$
 \caption{Consensus-based Distributed Variational Multi-object Tracker at time step $n$}
 \label{Algo:tracker}
\end{algorithm}
\section{Consensus-based Distributed Variational Multi-object Tracker}\label{sec distributed}
In this section, we present distributed variational filtering frameworks for a sensor network without a fusion center.
The aim is to achieve the same converged variational distribution $q_n^*(X_n,\theta_n)$, as obtained in the centralised variational filtering framework for approximating the target posterior $\hat{p}_n(X_{n},\theta_{n}|Y_n)$ in \eqref{eq:phatjoint}, by solely relying on local processing and communications between neighbouring sensors.
To this end, we assume at the initial time step $0$, an identical target state prior $p({X}_0)$ is given to all sensors $s=1,...,N_s$. 
To ensure the variational distribution at each sensor $s$ converges to the same $\hat{p}_n(X_{n},\theta_{n}|Y_n)$ at the time step $n$, according to \eqref{eq: X update}, it requires the local sensor has access to the pseudo-measurements $\overbar{Y}_{n}^k$ and $\overbar{R}_n^k$, $k=1,2,...,K$ calculated using all values of $\{\Omega_{k,1}^{s}, \Omega_{k,2}^{s}\}_{s=1}^{N_s}$ computed at each sensor. 

The sum expressions of \eqref{eq:pseudomeas Y} can be
computed at each sensor by using a distributed, iterative consensus algorithm. 
Specifically, we adopt the distributed average consensus algorithm introduced in \cite{xiao2005scheme}, which is guaranteed to converge provided that the sensor network is connected, even under time-varying communication links. For our application, given the initial value of $\Omega_{k,1}^{s}, \Omega_{k,2}^{s}$ at each sensor $s$, each sensor can converge to the same average value 
$\hat{\Omega}_{k,1}^{s}=\frac{1}{N_s}\sum_{s=1}^{N_s}\Omega_{k,1}^{s}$ and $\hat{\Omega}_{k,2}^{s}=\frac{1}{N_s}\sum_{s=1}^{N_s}\Omega_{k,2}^{s}$. As an example, the distributed average
consensus for computing $\hat{\Omega}_{k,1}^{s}$ at sensor $s$ can be described as follows.
\begin{itemize}
    \item At initial iteration $m=0$, each sensor node $s$ initialises its state as $\hat{\Omega}_{k,1}^{(s,0)} = \Omega_{k,1}^{s}$. 
    \item For $m=0,1,2,...$ until convergence\\
          each sensor $s$ updates its state by using its own state and the states of instantaneous neighbours $\mathcal{N}_s(m)$:
\begin{equation}
 \hat{\Omega}_{k,1}^{(s,m+1)} = W_{ss}^{(m)} \hat{\Omega}_{k,1}^{(s,m)} + \sum_{j \in \mathcal{N}_s(m)} W_{sj}^{(m)} \hat{\Omega}_{k,1}^{(j,m)} 
 \vspace{-0.5em}
\end{equation}
where $W_{sj}^{(m)}$ is the linear weight on $\hat{\Omega}_{k,1}^{(j,m)}$ at node $s$. Here we adopt the Metropolis weight in \cite{xiao2005scheme}:
\begin{equation}
W_{sj}^{(m)} =
\begin{cases}
\frac{1}{1 + \max{\{d_s^{(m)}, d_j^{(m)}\}}} & \text{if } j \in \mathcal{N}_s^{(m)}, \\
1 - \sum_{{s,k} \in \mathcal{E}(m)} W_{sk}^{(m)} & \text{if } j = s 
\end{cases}
\end{equation}
\end{itemize}
In the same way, we can obtain the $\hat{\Omega}_{k,2}^{s}$ by the same distributed average
consensus algorithm.
After obtained the converged value of $\hat{\Omega}_{k,1}^{s}$ and $\hat{\Omega}_{k,2}^{s}$, at each sensor $s$, we can compute the required pseudo-measurements $\overbar{Y}_{n}^k$ and $\overbar{R}_n^k$, $k=1,2,...,K$, by the following expressions:
\begin{align}\label{eq:average consesus result}
   \overbar{R}_n^k={( N_s \hat{\Omega}_{k,1}^{s})}^{-1},~~~~   \overbar{Y}_{n}^k = \overbar{R}_n^k ( N_s \hat{\Omega}_{k,2}^{s})
\end{align}
In this way, during each iteration of the update for $q_n({X}_n)$, every sensor $s$ updates $q_n({X}_n)$ locally based on the centralised pseudo-measurements calculated at all sensors using the distributed consensus algorithm, such that each sensor can behave equivalently to the fusion center in the centralised version.
The overall distributed implementation of the variational tracker is summarised in Algorithm \ref{Algo:tracker}.

\begin{figure}[tp!]
\centering    
\includegraphics[width=6.5cm]{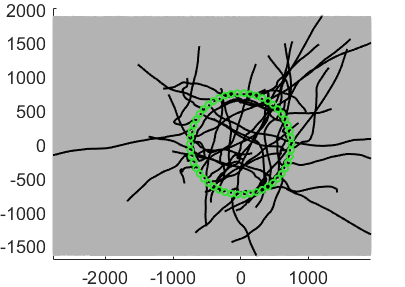}
\caption{Measurements, ground truth tracks of 50 targets; grey dots are measurements covering the whole background, black lines are the trajectories and green circles are starting points}
\label{simutrack}
\end{figure}

\section{Results}
In this section, we will conduct a performance comparison between different versions of the variational multi-object tracker: the centralised fusion in Section \ref{sec: cavf}, the optimal distributed fusion in Section \ref{sec distributed}, and the suboptimal distributed fusion with an arithmetic average (AA) fusion strategy. Specifically, the suboptimal distributed fusion adopts the similar approximation in the literature, e.g., \cite{li2021distributed}, in which each sensor infers a multi-object posterior distribution based on local measurements and, then a distributed average consensus algorithm is implemented to fuse the derived multi-object posteriors from each sensors using the AA fusion principle. 

\begin{figure}[tp!]
\centering    
\includegraphics[width=5.5cm]{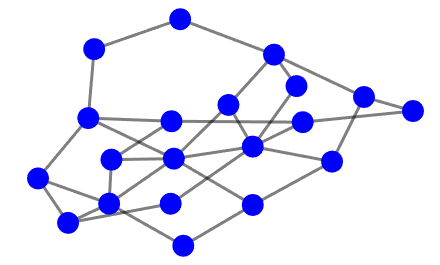}
\caption{Simulated sensor network; blue circles are sensors and lines indicate the communication links between sensors}
\label{sensornetwork}
\end{figure}

In the simulated dataset, the network consists of 20 sensors as shown in Fig. \ref{sensornetwork}, all observing 50 targets in the surveillance area. The general parameter settings are as follows. For all datasets, the total time steps are 50, and the time interval between observations is $\tau=1$s. The parameters in the CV model are $\sigma_{k}=5$ and $R_k^s=100\text{I}$ where I is a $2$-D identity matrix. The target Poisson rates are set to 1; the clutter rate is 100. To evaluate the robustness of the algorithm, we generate 20 different measurement sets under the same parameter settings. One sample measurement set is shown in Figure \ref{simutrack}. 
We use the optimal sub pattern assignment (OSPA) \cite{schuhmacher2008consistent} metric to evaluate the tracking performance of all methods. For the OSPA metric, the order is set to $p=1$ and the distance cut-off value is $c=50$. For both datasets, we calculate the mean OSPA metric over all the sensors and Monte Carlo runs. 

Figure \ref{ospa} shows the three variational multi-target trackers' mean OSPA of each time step calculated over all the sensors and Monte Carlo runs. Specifically, for both proposed optimal distributed fusion and the suboptimal AA fusion, we set the average consensus iteration to 20 to obtain the results in Figure \ref{ospa}. It is observed that the proposed optimal distributed fusion has a much lower mean OSPA value compared to the suboptimal AA fusion. The estimation results also confirm the equivalence of our proposed optimal distributed variational tracker with the centralised variational tracker when the distributed average consensus reaches convergence.
Figure \ref{ospa_iter} shows mean OSPA over all the sensors, Monte Carlo runs, and 50 time steps versus the number of iterations used in the distributed average consensus algorithm for proposed optimal distributed variational tracker. We can see that as the number of iterations increases, the performance of the optimal distributed fusion approaches the performance of the centralised fusion within approximately 10 iterations.

\begin{figure}[tp!]
\centering    
\includegraphics[width=8cm]{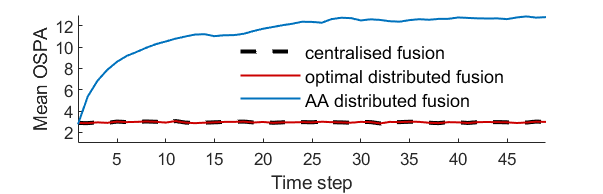}
\caption{Mean OSPA of different fusion strategies over 50 time steps, averaged over 20 Monte Carlo runs (average consensus iteration for both proposed optimal distributed fusion and the AA distributed fusion is 20)}
\label{ospa}
\vspace{-1em}
\end{figure}

\begin{figure}[tp!]
\centering    
\includegraphics[width=7.5cm]{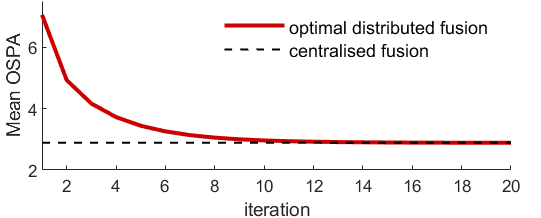}
\caption{Mean OSPA of the optimal distributed fusion over different iterations of average consensus algorithm, averaged over 20 Monte Carlo runs}
\label{ospa_iter}
\vspace{-1em}
\end{figure}

\section{Conclusion}
The paper presents a novel optimal distributed variational multi-target tracker for sensor networks that only require communication between neighbouring sensors. Our method achieves equivalent tracking performance to centralised fusion while retaining a decentralised processing architecture and reducing communication costs. The simulation results demonstrate the equivalence of the proposed optimal distributed fusion and the centralised fusion in terms of tracking accuracy. In the future, we will extend the current distributed variational tracker to handle unknown target numbers and heterogeneous sensor networks with varying coverage. 

\section{Acknowledgements}
This research is sponsored by the US Army Research Laboratory and the UK MOD University Defence Research Collaboration (UDRC) in Signal Processing under the SIGNeTS project. It is accomplished under Cooperative Agreement Number W911NF-20-2-0225. The views and conclusions in this document are of the authors and should not be interpreted as representing the official policies, either expressed or implied, of the Army Research Laboratory, the MOD, the U.S. Government or the U.K. Government. The U.S. Government and U.K. Government are authorised to reproduce and distribute reprints for Government purposes notwithstanding any copyright notation herein.



\bibliographystyle{IEEEtran}
\bibliography{./IEEEabrv,IEEEexample}

\end{document}